\documentclass[%
 aps,prl,showpacs,superscriptaddress,altaffilletter,numerical,amsmath,amssymb,floatfix,
reprint,longbibliography%
]{revtex4-2}

\usepackage{bm}%
\usepackage{graphicx}
\usepackage{}
\usepackage{marginnote}
\usepackage{xcolor}
\begin{document}

\title{Robust formation of nanoscale magnetic skyrmions in easy-plane thin film multilayers with low damping}%

\author{Luis Flacke}
\email{luis.flacke@wmi.badw.de}
\affiliation{\mbox{Walther-Meißner Institut, Bayerische Akademie der Wissenschaften, 85748 Garching, Germany}}
\affiliation{Physics Department, Technical University of Munich, 85748 Garching, Germany}

\author{Valentin Ahrens}
\affiliation{Department of Electrical and Computer Engineering, Technical University of Munich, 80333 Munich, Germany}

\author{Simon Mendisch}
\affiliation{Department of Electrical and Computer Engineering, Technical University of Munich, 80333 Munich, Germany}

\author{Lukas Körber}
\affiliation{Helmholtz-Zentrum Dresden-Rossendorf e.V., Institute of Ion Beam Physics and Materials Research, 01328 Dresden, Germany}
\affiliation{Fakultät Physik, Technische Universität Dresden, 01062 Dresden, Germany}

\author{Tobias Böttcher}
\affiliation{Fachbereich Physik and Landesforschungszentrum OPTIMAS, Technische Universität Kaiserslautern, 67663 Kaiserslautern, Germany}

\author{Elisabeth Meidinger}
\affiliation{\mbox{Walther-Meißner Institut, Bayerische Akademie der Wissenschaften, 85748 Garching, Germany}}
\affiliation{Physics Department, Technical University of Munich, 85748 Garching, Germany}

\author{Misbah Yaqoob}
\affiliation{\mbox{Walther-Meißner Institut, Bayerische Akademie der Wissenschaften, 85748 Garching, Germany}}
\affiliation{Physics Department, Technical University of Munich, 85748 Garching, Germany}

\author{Manuel Müller}
\affiliation{\mbox{Walther-Meißner Institut, Bayerische Akademie der Wissenschaften, 85748 Garching, Germany}}
\affiliation{Physics Department, Technical University of Munich, 85748 Garching, Germany}

\author{Lukas Liensberger}
\affiliation{\mbox{Walther-Meißner Institut, Bayerische Akademie der Wissenschaften, 85748 Garching, Germany}}
\affiliation{Physics Department, Technical University of Munich, 85748 Garching, Germany}

\author{Attila Kákay}
\affiliation{Helmholtz-Zentrum Dresden-Rossendorf e.V., Institute of Ion Beam Physics and Materials Research, 01328 Dresden, Germany}

\author{Markus Becherer}
\affiliation{Department of Electrical and Computer Engineering, Technical University of Munich, 80333 Munich, Germany}

\author{Philipp Pirro}
\affiliation{Fachbereich Physik and Landesforschungszentrum OPTIMAS, Technische Universität Kaiserslautern, 67663 Kaiserslautern, Germany}

\author{Matthias Althammer}
\affiliation{\mbox{Walther-Meißner Institut, Bayerische Akademie der Wissenschaften, 85748 Garching, Germany}}
\affiliation{Physics Department, Technical University of Munich, 85748 Garching, Germany}

\author{Stephan Geprägs}
\affiliation{\mbox{Walther-Meißner Institut, Bayerische Akademie der Wissenschaften, 85748 Garching, Germany}}

\author{Hans Huebl}
\affiliation{\mbox{Walther-Meißner Institut, Bayerische Akademie der Wissenschaften, 85748 Garching, Germany}}
\affiliation{Physics Department, Technical University of Munich, 85748 Garching, Germany}
\affiliation{\mbox{Munich Center for Quantum Science and Technology (MCQST), 80799 Munich, Germany}}

\author{Rudolf Gross}
\affiliation{\mbox{Walther-Meißner Institut, Bayerische Akademie der Wissenschaften, 85748 Garching, Germany}}
\affiliation{Physics Department, Technical University of Munich, 85748 Garching, Germany}
\affiliation{\mbox{Munich Center for Quantum Science and Technology (MCQST), 80799 Munich, Germany}}

\author{Mathias Weiler}
\email{weiler@physik.uni-kl.de}
\affiliation{\mbox{Walther-Meißner Institut, Bayerische Akademie der Wissenschaften, 85748 Garching, Germany}}
\affiliation{Physics Department, Technical University of Munich, 85748 Garching, Germany}
\affiliation{Fachbereich Physik and Landesforschungszentrum OPTIMAS, Technische Universität Kaiserslautern, 67663 Kaiserslautern, Germany}


\begin{abstract}
	We experimentally demonstrate the formation of room-temperature skyrmions with radii of about 25\,nm in easy-plane anisotropy multilayers with interfacial Dzyaloshinskii-Moriya interaction (DMI). We detect the formation of individual magnetic skyrmions by magnetic force microscopy and find that the skyrmions are stable in out-of-plane fields up to about 200\,mT. We determine the interlayer exchange coupling as well as the strength of the interfacial DMI. Additionally, we investigate the dynamic microwave spin excitations by broadband magnetic resonance spectroscopy. From the uniform Kittel mode we determine the magnetic anisotropy and low damping $\alpha_{\mathrm{G}} < 0.04$. We also find clear magnetic resonance signatures in the non-uniform (skyrmion) state. Our findings demonstrate that skyrmions in easy-plane multilayers are promising for spin-dynamical applications.
\end{abstract}
\date{February 2021}%
\revised{February 2021}%

\maketitle

Magnetic skyrmions~\cite{Muhlbauer2009} attract increasing interest for their potential use in novel devices for information storage and processing. Atomic-scale skyrmion lattices~\cite{Heinze2011} and low threshold current densities for spin transfer torque movement of skyrmions were reported~\cite{Jonietz2011}. The room-temperature stabilization of skyrmions in entirely metallic, thin-film heterostructures~\cite{Moreau-Luchaire2016,Boulle2016} has been identified as an important step towards real-world applications in information technologies.
However, uniting all properties required for applications in a single material system is challenging.
Metallic multilayer systems can be fabricated by sputter deposition techniques and allow to stabilize skyrmions at room temperature by the interface Dzyaloshinskii-Moriya interaction (iDMI)~\cite{Moreau-Luchaire2016, Woo2016}. This makes them promising candidates to resolve the challenges imposed by material design. For thin-film systems, an uniaxial anisotropy has been suggested to be a requirement for skyrmion formation in bulk-DMI~\cite{Wilson2014} and iDMI~\cite{Wang2018} systems, which translates to perpendicular magnetic anisotropy (PMA) in thin-film heterostructures. However, both the large spin-orbit coupling materials required to induce iDMI and the very small thickness of the ferromagnet necessary for the PMA, increase damping of the magnetization dynamics due to spin pumping~\cite{Tserkovnyak2002}. In turn, the high damping leads to smaller skyrmion velocities or, alternatively, requires higher current densities for fast skyrmion motion~\cite{Iwasaki2013a,Fert2017}. In addition, the PMA multilayer skyrmion systems usually is associated with a low field stability of skyrmions, as the spin textures shrink with increasing magnetic field and are annihilated by rather small external fields of $10$\,mT\,-\,$100$\,mT~\cite{Moreau-Luchaire2016, Boulle2016, Woo2016}.
The two challenges, high magnetic damping and low field stability can be overcome as discussed recently by Banerjee \textit{et al.}~\cite{Banerjee2014}. They show that switching from uniaxial to a two-dimensional (2D) easy plane anisotropy allows to significantly enhance the field robustness of the skyrmions. With this lifted PMA restriction, it is thus possible to use thicker ferromagnetic (FM) layers which still support sufficiently iDMI to create skyrmions but result in lower damping due to spin pumping.

\begin{figure*}[t]
	\centering
	\includegraphics{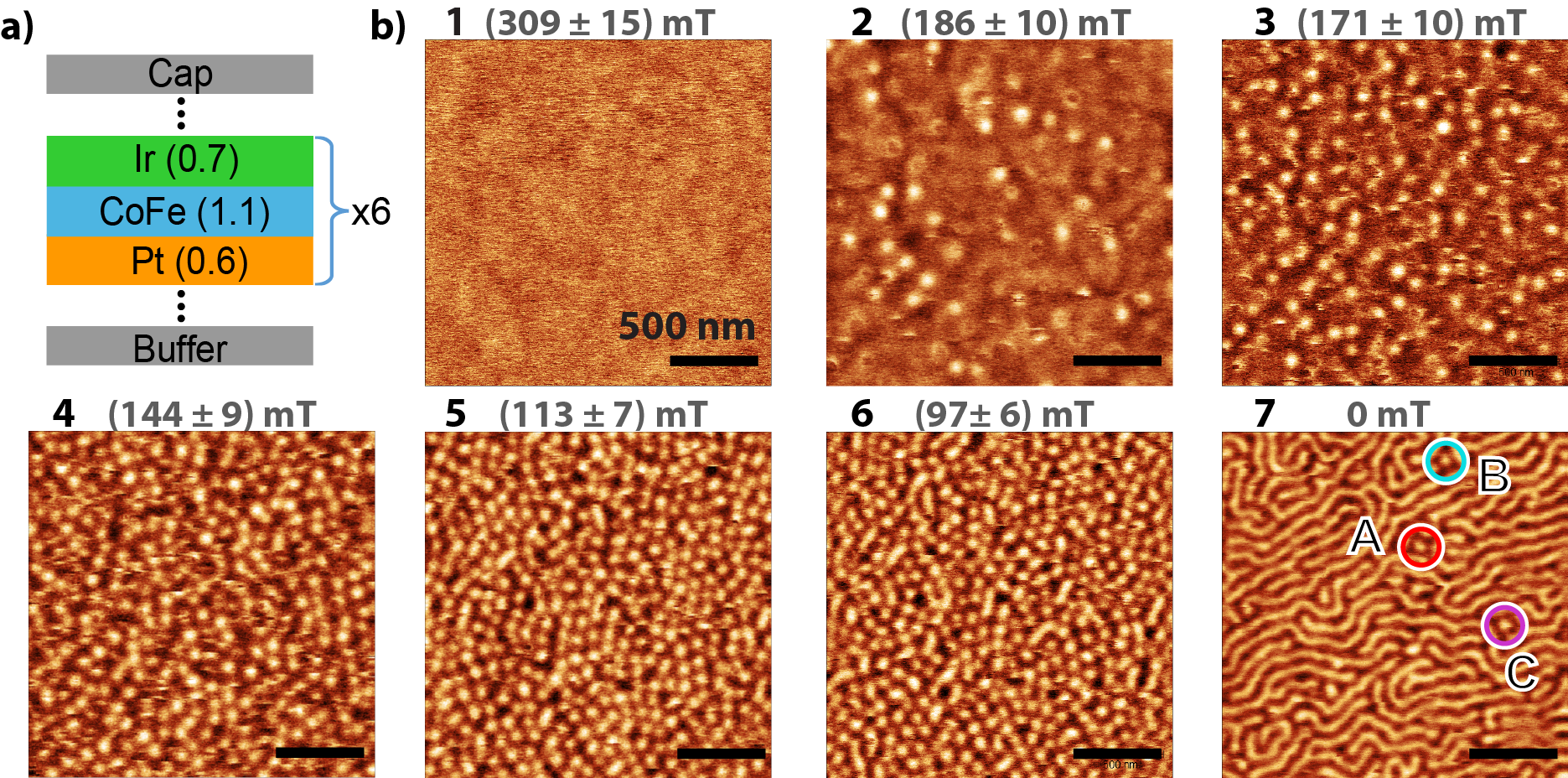}
	\caption{a) Sketch of the multilayer system, with a sixfold repetition of the Pt/CoFe/Ir trilayer (numbers give the film thickness in nm). b) MFM images recorded with OOP magnetic fields, field values are indicated at the top of each frame. With decreasing field, the magnetic texture shows transitions from a saturated ferromagnetic state ($1$), to isolated skyrmions ($2,3$), to a dense, skyrmion arrangement ($4-6$) and eventually forms a maze state at zero field ($7$). Individual skyrmions between the spin-spirals can still be found at zero field. The three highlighted skyrmions (marked with three colored circles in ($7$)) are analyzed further in Fig.~\ref{Fig-rad}\,a). Scalebar in all images corresponds to 500\,nm.}
	\label{Fig-1}
\end{figure*}

In this work, we experimentally explore the use of a 2D easy plane anisotropy in magnetic multilayers for the stabilization of skyrmions. The multilayer system based on Pt, CoFe and Ir also establishes the required iDMI and hosts skyrmions with diameter of less than 100\,nm while showing comparatively low damping of their magnetization dynamics.

Small skyrmions in thin-film multilayers can be stabilized by the iDMI, which favors perpendicularly aligned neighboring spins $\textbf{S}_\mathrm{1,2}$:
\begin{equation*}
	H_\mathrm{DMI} = - \textbf{D}_\mathrm{12} \cdot \left( \textbf{S}_\mathrm{1} \times \textbf{S}_\mathrm{2} \right)
\end{equation*}
The sign and magnitude of the iDMI vector $\textbf{D}_\mathrm{12}$ is determined by the involved materials and their stack sequence, allowing for additive DMI contributions from the two opposite interfaces of a magnetic thin film~\cite{Moreau-Luchaire2016,Woo2016,Boulle2016,Soumyanarayanan2017,Yang2018}. 

Thin-film multilayer systems employing iDMI for skyrmion formation typically rely on PMA. This is reflected in the condition for the effective anisotropy $K_\mathrm{eff} = K_\mathrm{u} - \mu_\mathrm{0}M_\mathbf{s}^{2}/2 > 0$, which omits higher order spin orbit coupling terms and dipolar contributions~\cite{Wilson2014, Wang2018}. Here, $K_\mathrm{u} >0 $ represents an anisotropy with easy axis along the film normal (out-of-plane, OOP), and the second term describes the shape anisotropy of the thin film field with the vacuum permeability $\mu_\mathrm{0}$ and the saturation magnetization $M_\mathrm{s}$. Recent theoretical calculations however also predict the existence of skyrmions with increased field stability within the $K_\mathrm{eff}<0$ regime~\cite{Banerjee2014,Lin2015,Rowland2016,Vousden2016}. Not requiring PMA allows to use thicker ferromagnetic (FM) layers with reduced spin pumping damping contributions, while iDMI is still sufficiently large to enable skyrmion formation.

We use sputter deposited heterostructues composed of $\mathrm{[Pt(0.6)/Co_{25}Fe_{75}(1.1)/Ir(0.7)]_{6}}$ multilayers (numbers give the layer thickness in nanometers), as sketched in Fig.~\ref{Fig-1}\,a) to reveal the abovementioned properties and benefits. Additionally, we fabricated a sample series with varying Ir thickness in order to determine the interlayer RKKY coupling. In the following, we denote the specific $\mathrm{Co_{25}Fe_{75}}$ alloy simply as CoFe. The details of the sample fabrications are discussed in the supplementary information (SI)~\cite{SupInf}. 

In order to observe of the formation of skyrmions in our multilayers, we use magnetic force microscopy (MFM). Fig.~\ref{Fig-1}\,b) presents $2\,\mathrm{\mu m} \times 2\,\mathrm{\mu m}$ MFM phase contrast images recorded with external magnetic fields $0\,\mathrm{mT} \leq \mu_\mathrm{0}H_\mathrm{ext} \leq 309\,\mathrm{mT}$ applied in the OOP direction.
In the following, we discuss the individual MFM images shown in Fig.~\ref{Fig-1}\,b). MFM image ($1$) shows vanishing contrast, as the sample is magnetically saturated, meaning that all magnetic moments are aligned parallel to the applied magnetic field. By reducing $\mu_\mathrm{0}H_\mathrm{ext}$ below the saturation magnetic field, small individual dots with an apparent diameter below 100\,nm start to emerge ($2$), which we attribute to skyrmion formation. A fixed chirality of the skyrmions is expected, as we determine a high interface DMI value of $D_\mathrm{int} \approx 1.86\,\mathrm{mJ/m^2}$ with BLS measurements~\cite{Nembach2015} on reference samples (see SI~\cite{SupInf}). This value agrees well with DMI strengths reported for superlattices with similar materials~\cite{Soumyanarayanan2017}. The density of skyrmions increases further on reducing the magnetic field magnitude until a dense, unordered arrangement is formed ($3-6$). Even lower magnetic fields eventually lead to the formation of a labyrinthine state at remanence, shown in image ($7$). However, obviously individual skyrmions maintain stabilized even at $H_\mathrm{ext} = 0$, as indicated by the three circles. 
The multilayer thus hosts skyrmions for $0 \leq \mu_\mathrm{0}H_\mathrm{ext} \leq 190\,\mathrm{mT}$. This field range for skyrmion formation is roughly twice the range reported for to $K_\mathrm{eff} \geq 0$ systems~\cite{Soumyanarayanan2017, Moreau-Luchaire2016, Legrand2020, Woo2016, Zeissler2017}.

\begin{figure}[]
	\centering
	\includegraphics{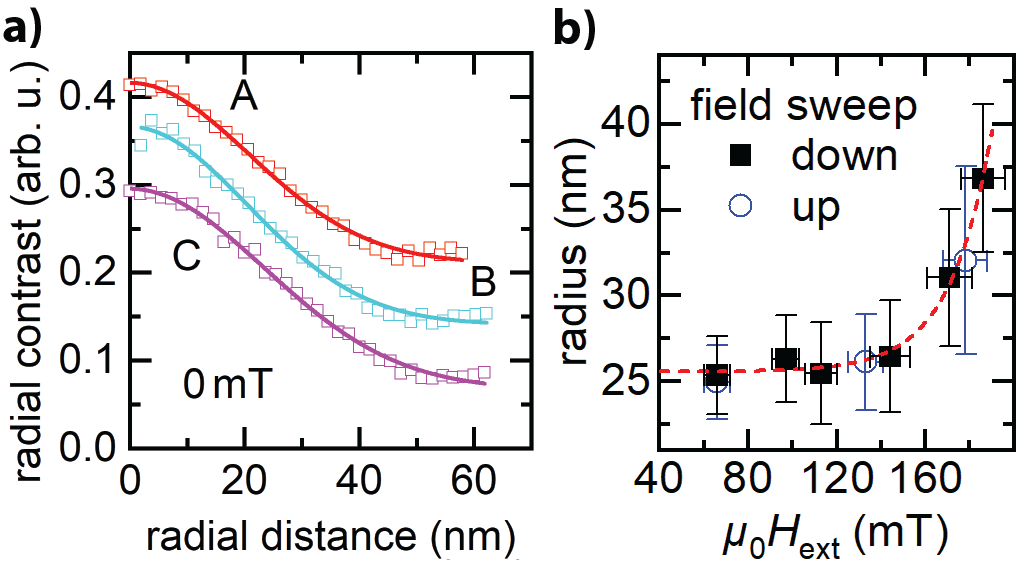}
	\caption{a) Azimuthally averaged MFM phase contrast of the skyrmions highlighted in Fig.~\ref{Fig-1} (offset is used for clarity). The radial profile is fitted with a Gaussian function, from which the radius $r = (25.4 \pm 1.6)\,\mathrm{nm}$ is found at $H_\mathrm{ext} =0$. b) Apparent radius vs. applied magnetic field in OOP direction. The black squares indicate the average value taken from 28\,-\,115 skyrmions from images ($2-6$) of Fig.~\ref{Fig-1}\,b) for a field down sweep starting at saturation, whereas, the blue circles indicate skyrmion sizes determined from an up sweep starting at $(66\pm6)$\,mT. The $y$-error represents the standard deviation of the size distribution. The $x$-error bar indicates the uncertainty of the field calibration. The red dotted line is a guide to the eye.}
	\label{Fig-rad}
\end{figure}

In Fig.\,~\ref{Fig-rad} we take a closer look at the magnetic field dependent skyrmion size distribution to elucidate the qualitatively different behavior compared to PMA systems~\cite{Moreau-Luchaire2016, Boulle2016, Romming2015}. Fig.~\ref{Fig-rad}\,a) depicts the azimuthally averaged radial profile of the magnetic contrast of the three highlighted skyrmions in image ($7$) of Fig.\,\ref{Fig-1}\,b) vs. the radial distance from their center position. Using a Gaussian fit, we extract the skyrmion diameter $d$ as the full-width-half-maximum of the MFM signal. We define the apparent radius as $r=d/2$. At $\mu_\mathrm{0}H_\mathrm{ext}= 0$\,mT, the three extracted radii are about 25\,nm. For larger $\mu_\mathrm{0}H_\mathrm{ext}$ we extracted the radial profile of 28\,-\,115 skyrmions from graphs ($2-6$) and plot the average radius and its standard deviation vs. the applied field. For $\mu_\mathrm{0}H_\mathrm{ext} < 150$\,mT, $r$ remains unchanged within experimental uncertainty, while $r$ diverges at the transition between the skyrmion state and the ferromagnetic alignment. This behavior is in contrast to the typically observed reduction of $r$ with $H_\mathrm{ext}$ observed for skyrmions in PMA-based multilayers~\cite{Moreau-Luchaire2016, Soumyanarayanan2017, Boulle2016}. We additionally verified experimentally that the skyrmion formation shown in Fig.~\ref{Fig-1} is non-hysteretic within experimental uncertainty, i.e., the same skyrmion size is obtained for given $\mu_\mathrm{0}H_\mathrm{ext}$, regardless of field history (see SI~\cite{SupInf}). This indicates a strong influence of the iDMI on the skyrmion formation in our samples.

\begin{figure}[h]
	\centering
	\includegraphics{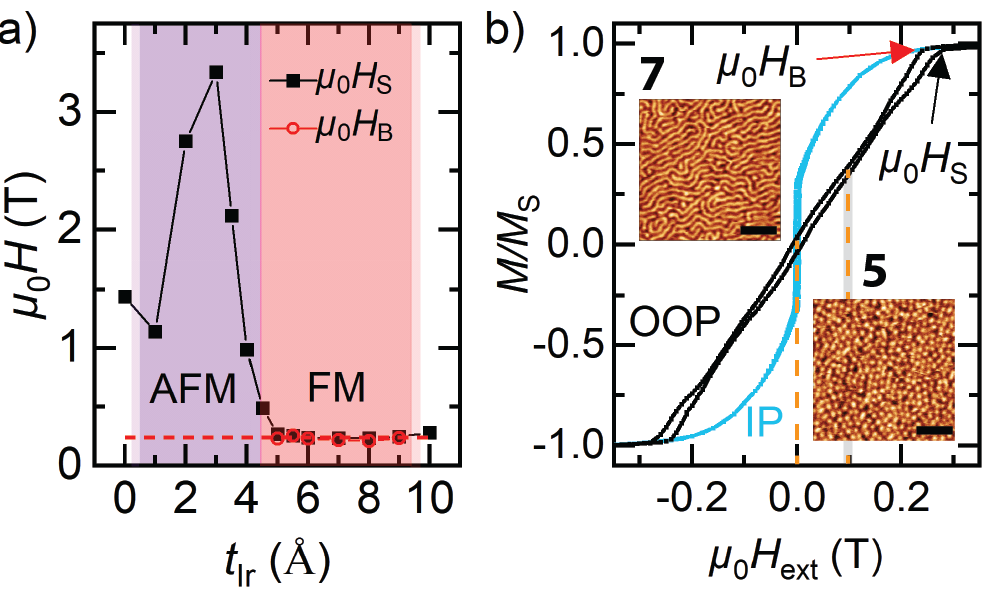}
	\caption{a) Saturation fields $\mu_\mathrm{0}H_\mathrm{S}$ of samples with varying Ir thickness $t_\mathrm{Ir}$ are determined by anomalous Hall effect measurements. In the AFM RKKY coupling regime, $H_\mathrm{S}$ correlates with the strength of exchange coupling. Within the FM regime, $H_\mathrm{S}$ is independent of coupling strength, and small pockets in the hysteresis curves allow for a qualitative separation of the regimes with AFM and FM coupling. Corresponding samples show a characteristic field $\mu_\mathrm{0}H_\mathrm{B} < \mu_\mathrm{0}H_\mathrm{S}$ below which saturation starts to break down. b) The hysteresis curves from the sample discussed in Fig.~\ref{Fig-1} are recorded by SQUID magnetometry in IP and OOP direction. The insets depict MFM images at the indicated fields (vertical dotted lines) along the descending OOP branch. The IP graph exhibits an easy-plane switching loop with a coercivity of less than 4\,mT. The decrease of magnetic moment for $|\mu_\mathrm{0}H_\mathrm{ext}| < 200\,\mathrm{mT}$ is attributed to domain formation due to iDMI. Scalebar in all images corresponds to 500\,nm.}
	\label{Fig-Squid-2}
\end{figure}

We note that a qualitatively similar radius increase can be found in antiferromagnetically coupled skyrmion bilayer systems~\cite{Potkina2020}. However, no antiferromagnetic (AFM) coupling was found in our sample. By employing anomalous Hall effect (AHE) measurements to extract the saturation field $\mu_\mathrm{0}H_\mathrm{S}$ (see SI~\cite{SupInf}) and varying the Ir spacer thickness as shown in Fig.~\ref{Fig-Squid-2}\,a) the AFM and FM RKKY coupling regions can be separated. Ir has been shown to induce strong AFM coupling between magnetic layers in the ultra-thin limit~\cite{Hellwig2007, Gabor2017}. As the coupling strength varies significantly with Ir thickness, so does the required field to overcome the AFM order and obtain saturation. Within the FM regime, $\mu_\mathrm{0}H_\mathrm{S}$ stays constant~\cite{Karayev2019}. The sample studied by MFM has an Ir thickness $t_\mathrm{Ir} = 7\,\mathrm{\AA}$ and is thus FM coupled. This rules out AFM coupling as the origin of the observed independence of skyrmion size on external magnetic field magnitude. 

Our observation is in accordance with the theoretical work by Banerjee \textit{et al.} who predicted a stable skyrmion size for a large field range for easy-plane anisotropy thin films. The contribution of the easy-plane "compass" anisotropy to the free energy stemming from higher order spin-orbit coupling terms~\cite{Banerjee2014} can possibly explain why we observe an almost magnetic field independent skyrmion size. This is further supported by the observation of a qualitatively different skyrmion size evolution in our micromagnetic simulations that do not include anisotropies caused by higher order spin-orbit coupling terms (see SI~\cite{SupInf})

For our system, the easy plane anisotropy is verified by superconducting quantum interference device (SQUID) magnetometry. We show the in-plane (IP) and the OOP magnetization curves in Fig.\,\ref{Fig-Squid-2}\,b). The shape of the OOP $M$ vs. $H_\mathrm{ext}$ loop is characteristic for samples with a maze-state formation, where skyrmions are often found~\cite{Hellwig2007, Woo2016, Soumyanarayanan2017}. Due to domain formation, also PMA systems involving labyrinthine arrangement can exhibit a similar hysteresis curve~\cite{Hellwig2007, Soumyanarayanan2017}. We verify the easy plane anisotropy by performing IP SQUID magnetometry measurements. The IP behavior qualitatively differs from the OOP curve and also from $K_\mathrm{eff} \geq 0$ systems, where the IP curve shows hard-axis behavior~\cite{Soumyanarayanan2017, Legrand2020}. We observe an easy-plane switching loop with a coercivity of less than 4\,mT. The gradual decrease of magnetization for $\mu_\mathrm{0}H_\mathrm{ext} \leq 200\,\mathrm{mT}$ is attributed to domain formation. The formation of a maze state at remanence despite easy-plane anisotropy is attributed to the strong iDMI~\cite{Moon2019}.

\begin{figure}[]
	\centering
	\includegraphics{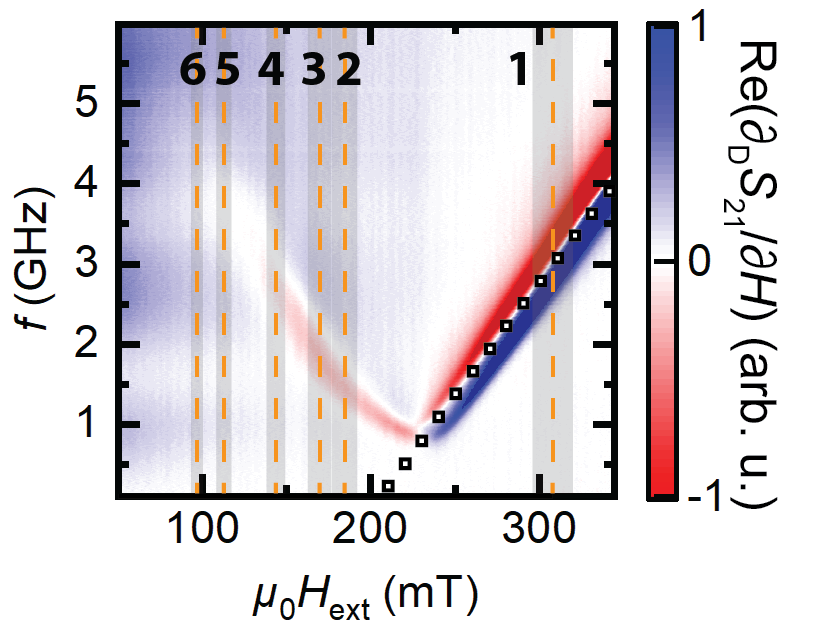}
	\caption{The colorplot shows the real part of the background corrected FMR signal. For $\mu_\mathrm{0}H_\mathrm{ext} > 240\,\mathrm{mT}$ applied along the OOP direction we see the characteristic FM response of the CoFe-multilayer. The black squares indicate the micromagnetic simulation results (see SI~\cite{SupInf}). Below the critical field we see an additional dynamic response, which we attribute to the skyrmion background. Dashed lines and the gray area indicate the field and its uncertainty, respectively, at which the correspondingly numbered MFM graphs in Fig.~\ref{Fig-1} were recorded. The Gilbert damping parameter is $\alpha_{\mathrm{G}} = (37 \pm 1)\times 10^{-3}$ (see SI~\cite{SupInf}).}
	\label{Fig-BBMR}
\end{figure}
We have so far demonstrated that stable skyrmions can form in easy-plane anisotropy thin films with DMI. We now turn to the dynamic magnetic properties of such thin film multilayers. To this end, we determine the dynamic microwave response of the sample by placing it on a coplanar waveguide (CPW) and recording the microwave transmission $S_\mathrm{21}$ through the CPW with a vector network analyzer (VNA)~\cite{Kalarickal2006, Weiler2017}. The frequency dependent background is removed by calculating the normalized field-derivative~\cite{Maier-Flaig2018} $\partial_\mathrm{D}S_{21}/\partial H$ and $
\mathrm{Re}(\partial_\mathrm{D}S_{21}/\partial H)$ is shown in Fig.~\ref{Fig-BBMR}.

For $\mu_\mathrm{0}H_\mathrm{ext} \gtrapprox  240$\,mT, we observe the ferromagnetic resonance (FMR) of the fully aligned CoFe magnetization, which follows the known OOP Kittel equation for thin films~\cite{Kittel1948}:
\begin{equation}
	2 \pi f_\mathrm{res} = \mu_\mathrm{0} \gamma (H_\mathrm{ext} - M_\mathrm{eff}).
	\label{Kittel}
\end{equation}
Here, $\gamma$ is the gyromagnetic ratio, $H_\mathrm{ext}$ is the applied magnetic field, and $ M_\mathrm{eff} = M_\mathrm{s} - H_\mathrm{k}$ with the perpendicular anisotropy field $H_\mathrm{k} = 2 K_\mathrm{u}/(\mu_\mathrm{0}M_\mathrm{S})$. We fit the data in Fig.~\ref{Fig-BBMR} to extract resonance fields and linewidths as detailed in the SI~\cite{SupInf}. We determine $\mu_\mathrm{0}M_\mathrm{eff} = (201.8 \pm 0.1)\,\mathrm{mT}$, again confirming the easy plane anisotropy. We used the determined parameters, as input variables for micromagnetic simulations and were able to qualitatively reproduce the hysteresis curves as well as the formation of chiral spin textures within the $\mu_\mathrm{0}M_\mathrm{eff} > 0$ system (see SI~\cite{SupInf}).
From fitting the linewidth of the signal for a broad frequency range we extract the Gilbert damping parameter and obtain $\alpha_{\mathrm{G}} = 0.037$ (see SI~\cite{SupInf}). This value is almost an order of magnitude lower than those reported on skyrmion host multilayers of Pt/CoFeB/MgO ($\alpha = 0.5$)~\cite{Litzius2017} and Ir/Fe/Co/Pt ($\alpha = 0.1$)~\cite{Soumyanarayanan2017}. Another common FM material used for skyrmion multilayers is Co, where $\alpha \approx 0.3$ is determined for layer thicknesses $t < 1\,\mathrm{nm}$ sandwiched between Pt layers~\cite{Metaxas2007}. As pointed out by Fert \textit{et al.}, such a decrease in magnetic damping is expected to result in substantial improvement for skyrmion motion~\cite{Fert2017}.

For $\mu_\mathrm{0}H_\mathrm{ext} < 240\,\mathrm{mT}$, $f_\mathrm{res}$ vs. $H_\mathrm{ext}$ shows a qualitatively different behavior with $f_\mathrm{res}$ increasing for decreasing $|H_\mathrm{ext}|$. This is attributed to a rotation of $M$ towards the film plane and the formation of a non-uniform magnetic texture in accordance with our MFM and SQUID magnetometry data. We attribute the resonance observed for $\mu_\mathrm{0}H_\mathrm{ext} < 240\,\mathrm{mT}$ to the dynamic precession of the magnetic moments in the quasi-uniform background (see Fig.~\ref{Fig-1}) and not to spin dynamics within the skyrmions. This is in accordance with the vanishing amplitude of the resonance towards $H_\mathrm{ext} = 0$, where no uniform background remains. Our observation is qualitatively different to that made by Montoya \textit{et al.}~\cite{Montoya2017a}, where various resonances of dipolar skyrmions in DMI-less Fe/Gd superlattices were observed. Even though the magnetic resonance data look similar, Montoya \textit{et al.} observed a phase transition of a field polarized state directly to a phase with coexisting stripes and skyrmions. Our MFM data, however, reveal a smooth transition from parallel alignment, over individual skyrmions, to a dense skyrmion arrangement and eventually a maze state formation. Our results are also qualitatively different from a study of PMA iDMI thin-films with magnetic skyrmions~\cite{Satywali2018}, where magnetic resonance is also observed in a phase with coexisting stripes and skyrmions.

We explicitly confirm a larger magnetic field range with skyrmion existence than comparable PMA systems~\cite{Soumyanarayanan2017, Moreau-Luchaire2016, Legrand2020, Woo2016, Zeissler2017} by MFM. Our measurements show a constant skyrmion size over a wide range of external magnetic fields (0\,mT to $\sim150$\,mT) as predicted by Banerjee \textit{et al.}~\cite{Banerjee2014}. This size evolution opposes the typical decrease of skyrmion size with increasing magnetic field due to Zeeman energy contributions~\cite{Moreau-Luchaire2016}. We explicitly rule out antiferromagnetic coupling between layers as an explanation for the unusual size dependence by varying the spacer thickness.
The easy-plane anisotropy in our films is verified by (SQUID) magnetometry and broadband magnetic resonance. Microwave spectroscopy experiments reveal a happily low damping within the system, making it promising for more detailed studies of spin dynamics. Compared to other experimentally realized easy-plane skyrmion systems like polar magnets~\cite{Bordacs2017}, bulk- and interface DMI thin films~\cite{Ahmed2018, He2018}, as well as DMI-less Fe/Gd superlattices~\cite{Montoya2017}, the investigated multilayer system uniquely provides room-temperature nanoscale skyrmions in a low damping host material for a large field range.
The significant reduction of damping can result in faster skyrmion motion~\cite{Fert2017}. The high stability of the skyrmion size over a broad range of external magnetic fields may be also beneficial for devices.

{\it Acknowledgments.} -- We acknowledge financial support by the Deutsche Forschungsgemeinschaft (DFG, German Research Foundation) via WE5386/4-1, WE5386/5-1 and Germany's Excellence Strategy EXC-2111-390814868.

\end{document}